# Effective team onboarding in Agile software development: techniques and goals

Jim Buchan
School of Engineering, Computer and
Mathematical Sciences
Auckland University of Technology
Auckland, New Zealand
jim.buchan@aut.ac.nz

Stephen G. MacDonell
School of Engineering, Computer and
Mathematical Sciences
Auckland University of Technology
Auckland, New Zealand
stephen.macdonell@aut.ac.nz

Jennifer Yang
School of Engineering, Computer and
Mathematical Sciences
Auckland University of Technology
Auckland, New Zealand
jennifer.hy.yang@gmail.com

*Abstract*— Context: It is not uncommon for a new team member to join an existing Agile software development team, even after development has started. This new team member faces a number of challenges before they are integrated into the team and can contribute productively to team progress. Ideally, each newcomer should be supported in this transition through an effective team onboarding program, although prior evidence suggests that this is challenging for many organisations. Objective: We seek to understand how Agile teams address the challenge of team onboarding in order to inform future onboarding design. Method: We conducted an interview survey of eleven participants from eight organisations to investigate what onboarding activities are common across Agile software development teams. We also identify common goals of onboarding from a synthesis of literature. A repertory grid instrument is used to map the contributions of onboarding techniques to onboarding goals. Results: Our study reveals that a broad range of team onboarding techniques, both formal and informal, are used in practice. It also shows that particular techniques that have high contributions to a given goal or set of goals. Conclusions: In presenting a set of onboarding goals to consider and an evidence-based mechanism for selecting techniques to achieve the desired goals it is expected that this study will contribute to better-informed onboarding design and planning. An increase in practitioner awareness of the options for supporting new team members is also an expected outcome.

## I. INTRODUCTION AND BACKGROUND

Much contemporary software development is undertaken by small cross-functional development teams, often using an Agile way of working [1]. A new team member may join an existing team, to replace a leaving team member or to augment the team's capacity. It is usual that the new team member will take some time to be integrated into the team as a productive and trusted contributor. This period of integrating and learning is referred to as "onboarding". We adapt a definition of organisational onboarding [2] to define team onboarding as: 'a procedure whereby new employees move from being team outsiders to becoming team insiders'.

If software development organisations have many teams or are in a stage of high growth they will have to deal with this situation of onboarding new team members frequently, and rapid and effective onboarding can have a high impact on organisational capacity. The time it takes to onboard a new team member may vary depending on factors such as their work experience, attitude and skills (e.g. [2], [3]). The duration of onboarding may also depend on the support given to the new team member (referred to as the "onboarder" from now on) during the onboarding period. It is this support, and how to optimise it for a given context, that is the subject of this study. This research investigates the activities and techniques used during onboarding that are most effective in achieving desired onboarding goals or outcomes.

A poorly designed or executed onboarding experience can cause the new team member anxiety at their lack of team contribution and trust, as well as a reduction in overall team productivity. This begs the question of what constitutes an *effective onboarding experience*? This study takes the perspective that an effective onboarding experience achieves a set of goals or desired outcomes for the onboarder, and that the likelihood of the goals being achieved can be increased by explicitly engineering a set of activities and experiences for the onboarder during this time of integration. We refer to these activities and experiences as *onboarding techniques* and a planned set of techniques over a set duration as an *onboarding program design*. Our aim is to better support software team onboarding by providing an empirical basis for organisations to create onboarding program designs for given contexts. We are interested in understanding practitioners' experiences and views on the efficacy of current onboarding techniques, and to what degree they think each technique contributes to common onboarding goals. The intent is that this could then be used by an organisation to select a set of onboarding techniques (an onboarding program design) that have a high likelihood of achieving a desired set of onboarding goals (that represent the context of the onboarding). The context for the team onboarding investigation reported in this paper is focused on software development teams using an agile way of working. This is a common situation and so should resonate with many organizations. The agile teams in the interviewees were working in had 5 to 10 team members.

This paper reports the results of an interview survey in which we asked eleven software practitioners from eight organisations what onboarding techniques they had experience with and found valuable to their own team onboarding. We then asked each participant their views on which onboarding techniques are important in achieving different onboarding goals in a list we derived from a literature review. This information was collected using a Repertory Grid Technique (RGT) where the participants indicated on a printed grid their perception of the contribution of each of their onboarding techniques to each onboarding goal in our list.

RGT has its foundations in George Kelly's Personal Construct Theory (PCT) created in the 1950's. This takes a constructivist philosophical perspective that individuals construct knowledge about the world around them by interpreting what they observe [4]. The type of partial RG used here has been used in similar studies (e.g. [5]) and has an accepted place in software engineering research [6]. The RGT is used to map conceptual elements to related constructs about a phenomenon. In our study the participants use their work experience to build up a personal construct system that is used as evidence for the research conclusions. In-line with similar studies, the personal construct system relating onboarding techniques (*constructs* in RGT) to onboarding goals (*elements* in RGT) was captured using a Repertory Grid (RG) instrument of rows and columns for each participant. In a full RGT both elements and constructs are elicited from the participants. In our case, the elements are the expectations of the software community of the goals of onboarding as stated in the literature and so were fixed by the researcher.

It is expected that this study will contribute to an increase in practitioners' awareness of the options for onboarding support in software development teams by presenting a set of techniques others have found effective. In addition, the results of the study should contribute to better-informed onboarding design and planning by presenting a set of onboarding goals to consider and an evidence-based mechanism for selecting the techniques to achieve the desired goals. These contributions should result in more effective onboarding, happier onboarders and more productive teams. The study also contributes to the body of empirically-based evidence related to software development in small teams.

The next section summarizes related work to situate the research presented in this paper. Section III describes the research process including the literature research protocol, the interview protocols and participant selection. The findings are then presented in Section IV, followed by Discussion of these findings. Section VI presents the main conclusions and threats to validity are discussed in Section VII.

## II. RELATED RESEARCH

While there are a number of existing empirical studies of onboarding, they mainly consider the situation of onboarding a new employee to an organization, rather than onboarding a new team member to a team. Moreover, only a few studies consider onboarding in the software development context. Two highly cited examples are a case study of software team onboarding of new employees at Google [7], and a study of onboarding in software teams of new graduates at Microsoft [8]. The first of these used semi-structured interviews to provide insights into Google's then state-of-the-art (2009) onboarding process for new graduates and proposed benchmarks to achieve their desired goals of reducing isolation and enhancing collegiality of the new hires. This is similar to the findings of our study which indicates the importance of team and leadership support as well team socialization as onboarding techniques. In the Microsoft study [8], the problems of new graduates in onboarding are identified and the root cause of these is traced to their poor communications skills and social naivete. The use of mentoring, pair programming and legitimate peripheral participation as onboarding techniques was shown to alleviate these issues. These findings also align with our findings where practitioners emphasized the importance of these three techniques in onboarding.

Another study provides some insights into onboarding in virtual teams [9] and concludes that onboarding support was almost non-existent due to few opportunities for the team members to interact. They showed that very poor onboarding practices can lead to poor coordination, reduced trust, and conflict between team members. Our study focuses on onboarders who are co-located with their teams, but the outcomes of poor onboarding found in [9] are likely to still apply in our context which is a motivating factor for our study.

A number of studies consider the factors that influence the efficacy of onboarding, identifying employee characteristics such as proactive personalities [10], ability to develop social networks [11], and openness to new experiences [3] as being important. While these are more relevant to the recruitment process, they may inform the customization of an onboarding program to suit the specific characteristics of the onboarder. The characteristics of the onboarder and the possible effects on the onboarding design are outside the scope of our study.

Studies such as that of Bauer *et al.* [12] focus on identifying enablers and strategies for effective onboarding. Related to this is a body of onboarding research (e.g. [13-15]) that investigates the importance of "organizational insiders" such as mentors. Organizational insiders that were identified as important to onboarding in our study include mentors, team leaders and team members, as well as members of other development teams.

Other onboarding studies focus on understanding the outcomes that are expected from effective onboarding (i.e., the goals of the onboarding). Bauer and Erdogan [2] proposed four such outcomes: (1) high onboarder satisfaction, (2) high onboarder organizational commitment, (3) low turnover, and (4) better onboarder performance. These results align with a number of other onboarding studies (e.g. [3], [16], and [17]). These studies take the perspective of the organization or managers, whereas our study identified desired onboarding goals or outcomes related specifically to the domain of software development from the perspective of the onboarder.

A synthesis of onboarding literature is also reported by Bauer and Erdogan [2]. They develop a model of onboarding as an adjustment of the onboarder's clarity of their role, self-efficacy, acceptance by organizational insiders, and knowledge of the organizational culture. Such positive adjustments to the onboarder during onboarding are posited as resulting in their increased satisfaction, commitment and performance. They propose that factors such as onboarder characteristics, onboarder behaviors, and organizational effort can affect the duration of these adjustments to achieve the outcomes. Our research identified similar areas of onboarding adjustment goals and extends this to a higher level of granularity in the context of Agile software teams. We then investigate what onboarding techniques should be used to support the onboarders' adjustments in each of these areas.

## III. THE RESEARCH APPROACH

### A. Research objectives, aims and questions

Our overall research objective is to gain insights into software practitioners' perceptions of the efficacy of onboarding techniques in practice and in particular their views on which onboarding practices they identify as valuable and will effectively support one or more onboarding goals.

The main research aims to achieve this research objective are to identify: (1) a set of high-level onboarding goals (desired outcomes of the onboarding program), (2) a set of onboarding techniques that practitioners perceive as effective for onboarding, (3) practitioners' perceptions on the level of contribution of each onboarding technique to each onboarding goal, (4) practitioners' perceptions of the duration of onboarding.

The stated research aims are realised through answering the following research questions:

RQ1. *What set of onboarding goals are relevant to the context of software development teams?* A subset of these will represent the desired outcomes of an onboarding program from the organization's perspective.

RQ2. *What is a suitable set of contemporary onboarding techniques useful in team software development?* A subset of these will comprise the basis of an onboarding program design.

RQ3. *Which onboarding techniques contribute strongly to the achievement of each onboarding goal?* This will provide a basis for selecting which onboarding techniques to implement for a desired set of onboarding goals.

RQ4. *What is the likely duration of onboarding?* The end point of onboarding is often uncertain. Some idea of the expected duration will assist with aligning expectations and planning.

An overview of the research process employed to answer the first three questions is presented in Fig 1, indicating which parts of the process addressed which research questions. The answer to RQ4 was based on participants' answers to the final interview question: "thinking back on your previous experience of joining a development team, how long do you think it took you before you felt you were a productive part of the team and were onboarded?".

### B. Literature Search (RQ1)

Three online research databases (Google Scholar, ACM Digital Library, and IEEE Xplore) were searched for relevant articles for the years 2007 to 2017. The search string was customized for each database but included keywords (onboarding OR newcomer OR new employee OR new team member OR new staff) AND (software OR software develop OR software engineer). This query returned 202 candidate articles. After excluding duplicates and any that were not related to onboarding, or were not empirical research, 20 papers were available for further analysis. In these papers any text that mentioned expected onboarding outcomes, areas of onboarder adjustment, challenges faced by onboarders, or onboarding goals were extracted as candidate onboarding goals for our RGT study. The relevant text extracts from the selected papers were read and categorized into conceptual areas of onboarding goals using a simple open coding technique following the steps in [18]. These concepts were then organized into themes by identifying recurring patterns. Eleven different areas and five themes for onboarding goals were identified through this process (see Table 1). Two of the authors independently coded and categorized the material and then discussed discrepancies to get consensus on the main desired onboarding goals or outcomes for a specific to the context of an Agile software development team. These goals in the second column of Table 1 were used as the elements (column headings) in the Repertory Grid and explained to participants as part of the interview protocol.

### C. Participant Selection

Participants from fifteen different organisations were approached from the personal networks of the authors. They were invited based on (1) the proximity of the organisation to the researchers (Auckland-based), (2) evidence that software development was a key aspect of the organisation (3) claim to develop software in small teams using an Agile approach (4) their willingness to participate. A total of fourteen participants from ten organisations initially accepted, but three did not go ahead.

The final eleven participants were from eight different organisations and included eight developers and three testers. Of these, seven were new to the organisation and the role, and five were recent graduates. Ten were new to working in an Agile way and all eleven were new to their software product domain. All participants drew on their own recent experiences of onboarding.

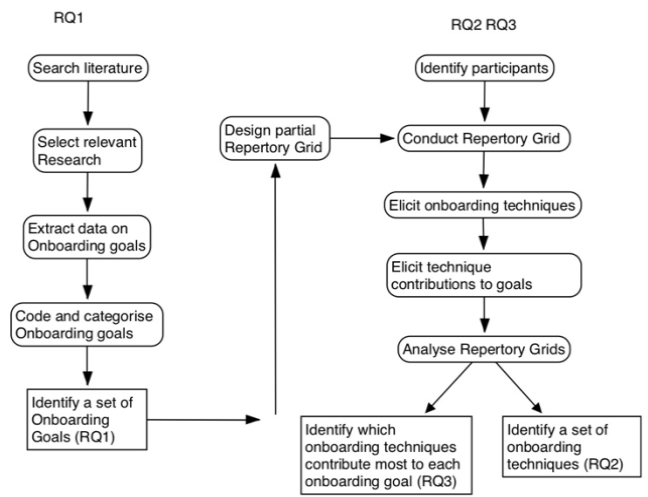

Fig. 1. Research Process

TABLE I. ONBOARDING GOALS

| Theme | Goal | Source References |
|---|---|---|
| Culture Context | Understand and fit in with company culture | [19] [20] [15] [21] [22] |
| | Understand and fit in with the team norms | [23] [24] [25] |
| Job Responsibility | Understand and meet others' expectations of one's own role's responsibilities | [26] [2] |
| | Understand the responsibilities, expertise and authority of other team members | [27] [23] [19] [25] |
| | Understand what work to do and when | [28] [29] |
| Standard of Work | Understand how to code and test to the team's expectations | [27] [30] |
| | Understand and meet the team's standards of work quality | [30] |
| Development Process | Understand and adopt the Agile mindset | [31] [32] |
| | Know how to use Agile artefacts and techniques used by the rest of the team | [33] [32] |
| Project Knowledge | Understand the short, medium and long term work structures, aims and implications | [22] |
| | Understand the product/project domain knowledge and terminology | [34] [26] |

The organisations represented by the participants were Small-Medium Enterprises (SMEs) with most having 20-100 employees, two with more than 200 and one with fewer than ten. Industry sectors represented were financial services, insurance, healthcare, telco, bespoke software development, fleet management, serving a mixture of internal and external clients. All participants claimed their teams developed software using the Scrum framework.

Prior to the interview all participants were provided with an information sheet outlining the research, its goals, what was expected of them and what they could expect of the researcher. Confidentiality and participant safety were assured.

*D. The Repertory Grid design*

The main components of RGT are (1) the *topic* or domain of the research (team onboarding); (2) the *elements,* which are instances of the topic that are the research objects (the onboarding goals that be achieved from onboarding experiencing the techniques ); (3) the *constructs,* which are the ideas the participants hold about the elements (onboarding techniques from recent onboarding experience ); (4) *the links,* which are the perceived relationships between the elements and constructs (what level of contribution does each onboarding technique make to each onboarding goal). The topic and elements (the onboarding goals) were pre-determined by the researcher in our case and the constructs and links were elicited from the participants. A 7-point Likert scale was used to elicit the level of contribution of each onboarding technique to the achievement of each goal: 1 represented no contribution and 7 represented a very high contribution, with 4 the mid-point. A 7-point scale was used to give sufficient discriminatory power to the participant but keep the effort manageable.

*E. The Interview Protocol*

The protocol involved interviewing one participant at a time with two researchers present at most interviews (three interviews were conducted by only one researcher for logistical reasons). The interviews were 45 to 60 minutes in duration and were generally conducted at a neutral place away from the participants' workplaces.

In the interview, after the initial administration tasks, we introduced the topic to the participant and asked them to recall their most recent team onboarding experience and describe what activities or experiences they had found valuable for their own onboarding. For each technique identified they were asked to give the onboarding technique a label or short description and explain how it was useful for their onboarding. The responses were recorded both as audio and in interview notes. Once the participant could think of no more techniques that they considered important to their onboarding, the list of technique labels was transposed from the field notes to the pre-prepared Repertory Grid form which had the onboarding goals as column headings and the onboarding techniques just elicited as row headings.

The participant was then asked to indicate what level of contribution (on the Likert scale) they felt each onboarding technique had for the achievement of each onboarding goal. They filled in each cell in the grid with a number from 1 to 7. Before starting this, each goal was explained to the participant using a set of standard definitions synthesised from literature. The participant could ask questions and get clarification of the goals anytime during the interview but was encouraged to get it clear before starting. The participants were asked to voice their thinking as they completed each grid to provide further insights. The researcher was careful not to influence the participant's thinking and generally remained silent.

*F. Data Analysis*

The RG data analysis involved quantifying the aggregated views of all participants. This amalgamated view represents the diversity and commonality of understanding of the phenomenon in the range of contexts and experiences of the participants. It is expected that this provides a richer baseline (compared to a single view) to support others to construct meaning of the phenomenon studied.

The first step in aggregating the RG data was to merge duplicates and synonyms from the lists of onboarding techniques elicited from each participant. This reduced the list from 81 to a set of 24 unique onboarding techniques. Each participant's repertory grid was then "normalised" by mapping the onboarding techniques they had identified to a subset of the 24 in the final list. This made it possible to identify patterns in the contributions of each technique to each onboarding goal by aggregating the scores of the individual cells in each RG.

The standard approach to aggregating RG data is to use the Frequency Distribution technique (e.g. [35], [36]). This involves treating the Likert scale as an interval scale rather than an ordinal scale and for each cell calculating the weighted average of the frequency distribution of the Likert values for that cell. To simplify the presentation of the results we converted the numerical aggregation results to five categories of contribution: Very High (VH), High (H), Medium (M), Low (L), and Very Low (VL) for each cell. A simple mapping was used, based on the numerical distance of an aggregated score from the middle Likert value (4): a score of more than one standard deviation (SD) below = VL; less than or equal to one SD below = L; the middle score = M; up to one SD above = H; more than one SD above = VH. This makes the common assumption that the aggregated Likert scores are normally distributed around the middle value.

The audio recordings were not transcribed in full but were referred to for clarification of field notes or filling in gaps in the field notes. The participants were asked to think out loud while completing the repertory grid so the audio is a rich data source, providing a detailed explanation of each technique.

A pilot of the interview and RGT was conducted with one participant, although this data were not included in the final results, since some changes to the questions and protocol were made based on the feedback and learning from this pilot.

IV. FINDINGS

Having searched the literature to identify the main goals of an onboarding program (Table 1), the next step is to identify the onboarding techniques that the interviewees considered significant to their own onboarding success. These are described in the next subsection. Following that, the results of aggregating the participants' views in the RGs are presented.

*A. Onboarding Techniques Used (RQ2)*

The reduced list of onboarding techniques identified by the interviewees is presented in Table II. They are ordered by the number of interviewees that identified them as important (n), as an indication of how 'front-of-mind' each technique was. The onboarding techniques are written from the perspective of the new team member as an activity that they identified as part of their own or others' recent onboarding experience. Even on its own, this list could be useful to practitioners as a checklist of possible ideas to adapt to their own specific context.

The onboarding techniques are also discussed in more detail in this section, based on the verbal explanations of the interviewees. This detail could be useful for onboarding program design and implementation since it provides insights into onboarders' perceptions of the meaning and value for each technique. Table II shows the onboarding techniques as categorized into three themes: (1) Working with people, (2) Working with artefacts and (3) Undertaking an activity. These themes are used to structure the detailed description of the onboarding techniques that follows.

TABLE II. ONBOARDING TECHNIQUES

| Technique | n | Description | Theme |
|---|---|---|---|
| Mentoring | 9 | Have an assigned mentor as an experienced person for regular and ad hoc face-to-face meetings and interactions. | People |
| Online Communities | 8 | Searching online communities such as Stack Overflow to find answers to specific technical questions. | Activity |
| Peer support | 7 | Ad hoc opportunities to ask peers (in and outside the team) for information or guidance. Usually face-to-face. This included observing others as they worked or met. | People |
| Team Socializing | 7 | Interacting with other team members in a social setting (not related to work tasks). | Activity |
| Training Course | 6 | Attend a formal course to achieve specific learning objectives or certification related to work. May involve availability of an "education stipend". Also includes online courses. | Activity |
| Review code | 6 | Analyze and understand relevant existing source code. Attend code reviews. Access to code repository. | Artefact |
| Internal documentation | 6 | Documentation capturing local knowledge about data structures, algorithms, and control flow of the project. May also include product information. | Artefact |
| Product overview | 5 | A presentation, video or similar that shows the functionality and features of their product as well as the business value. | Artefact |
| Pair program | 4 | Develop with another developer at one workstation, swapping between driving and navigating regularly. | Activity |
| Stand ups | 4 | Have regular team standup meetings as described in Scrum or adapted. | Activity |
| Simple task | 3 | Do task that is low risk and technically unchallenging, but provides experience with tools, process, technology, team norms. | Activity |
| Self-learning | 3 | Learn about libraries, tools and techniques with free access to books and online courses through sites such Lynda, Pluralsight, Udemy, Code Academy, MSDN | Activity |
| Induction | 3 | Learn about the company's history, beliefs, values, long term goals, and company structure as well as safety, security and health and job responsibilities, accountability and progression. | Activity |
| Knowledge database | 3 | Access and contribution to a local knowledge database such as a wiki may store complex structured and unstructured information. This may relate to product information, design decisions, testing architecture, coding standards. | Artefact |
| Team Leader support | 1 | Ad hoc assistance from the Team Leader (may be Scrum Master, or Project Manager) answering questions or explaining decisions. | People |
| Course on Agile | 1 | Learn about the Agile way of working by attending a course (usually third party but may be run by internal coaches). | Activity |
| Team retrospective | 1 | Review challenges and learning with the team and learn from their challenges and learning. Some teams do this as part of their sprint retrospective meetings. | Activity |
| Review plan | 1 | Review the longer term plan for the project to understand what has been done and is coming up. | Activity |
| Attend Conference | 1 | Learn from others by attending a relevant technical national, regional or international conference. | Activity |
| Set expectations | 1 | Expectations about onboarding activities and goals are explicitly discussed and set before onboarding and reviewed during onboarding. | Activity |
| Electronic communication | 1 | Get assistance from others in the organization through electronic communications such as email, chat, social media. | People |
| Meet with other teams | 1 | Face-to-face meetings with other teams in the organization. May be at different branches geographically separated. | Activity |
| Location map | 1 | A diagram showing the distribution of every staff member in the floor. The information of staff such as authority, expertise and department is also attached in the Floor map. | Artefact |
| Checklists | 1 | Given checklists to assist with remembering aspects of work. For example, a checklist of points to look for when reviewing others' code. | Artefact |

*1) Working with people*

*Mentoring* was mentioned as an onboarding technique experienced by all interviewees except for two from the same organization. The perception was that mentoring provided opportunities for both planned, regular feedback and guidance, as well as for *ad hoc* advice or assistance. The participants gave the sense that interactions with mentors could be initiated by either the mentor or mentee, i.e., as proactive or reactive assistance. The mentors tended to be experienced in the role in the organization, although two interviewees mentioned that their mentors had been in the job less than 6 months and themselves had limited experience. Senior developers or team leaders outside the team, experienced team members and Scrum Masters were all mentioned as mentors by the interviewees. Interviewees described good mentors as: trusting of the onboarder; trustworthy in their advice; trustworthy in confidentiality; nurturing rather than authoritarian; willing to be a mentor (saw value in it); available regularly as planned as well as *ad hoc* (had the time). Mentors were seen as someone to help technically at times, but also to give advice on people issues, how to work, and to give feedback on onboarding progress. It was seen as a key success factor of onboarding to have a mentor "*take them under their wing*". As one developer put it:

*In my opinion, mentoring is the most important way to help newcomers. For me, before my mentor came to me, I had no clue of nearly everything. I didn't know how to set up the working environment, who would I ask, and even where to start.*

For mentoring to be effective in onboarding, the organization has some responsibilities that were identified by interviewees as challenges they experienced: the mentor should have training in mentoring; the time taken for mentoring should be acknowledged in their job responsibilities; and expectations around the mentoring responsibilities should be explicitly agreed on by mentors and mentees.

The use of o*nline communities* in onboarding was front of mind for almost all interviewees. Interviewees said that often their first response to a technical problem while working on a task was to refer to online services such as Google search, Stack Overflow, Quora, Microsoft Developer Network (MSDN), experts' blogs. No participant mentioned how they judged the credibility of such sources of information and advice. This is working with people in the sense there is a set of electronic interactions (usually questions and answers) with virtual people resources. Mentoring is recognized as key in a number of onboarding studies. For example, [37] and [38] show that mentor-supported developers were more active earlier than developers with no mentoring during onboarding.

*Peer support* was also identified as important to onboarding by most interviewees. They described it as the support of other team members to answer questions, explain aspects of the work or assist with the actual work task. The team support tended to be reactive and self-driven, although one interviewee mentioned team support as including other team members making allowances for their "slow work and many questions". Another participant mentioned that they found it useful during onboarding to be "allowed" to observe others at work and participate in some meetings as an observer only (i.e., legitimate peripheral participation).

One of the challenges related to peer support mentioned in the interviews was the need to avoid interrupting other team members too often. Some interviewees reported that they preferred to use online/electronic resources to answer specific questions rather than asking other team members or mentors.

*2) tUndertaking Activities*

*Pair programming* was seen by some interviewees as an effective way of getting to know the existing code, how to code in the team context, as well as useful for getting to know other team members.

Another software development practice, the *daily stand-up* (scrum) meeting, was valued during onboarding by several interviewees as an opportunity for getting regular support from the other team members. It was seen as strengthening relationships as people talked about what they planned and their development problems. It was also a frequent and regular opportunity for the new team member to ask questions. One participant highlighted the onboarding value of the team meeting to reflect on how the team has been going and what they could learn (i.e., a *team retrospective*, although they did not use this terminology). The interviewee explained that during onboarding such a meeting improved their understanding of how the team thinks and works as well as providing a safe forum for them to have their say. The value of *meeting with other teams* was mentioned by one interviewee as important to onboarding by providing an opportunity to see what other teams did and potentially learn from this.

It is noteworthy that no interviewees mentioned that interacting with the client or Product Owner (PO) was an important aspect of onboarding, indicating this was not front-of-mind. It may be that relevant information from the client or PO (e.g., user requirements, business value, existing situation) was obtained as part of the team process and so no technique specific to onboarding was needed. Considering the values in the Agile Manifesto, an onboarding program should benefit from an explicit technique for the onboarder to understand the nature of the users, the client and the existing system.

Two interactions (with team leads or mentors generally) that were noted as particularly important for onboarding are *setting expectations* and *reviewing plans*. One interviewee described how important it had been to them to have their onboarding expectations discussed and progress reviewed periodically. This put their mind at ease, removing some of the uncertainty and speculation about how they should behave during the onboarding. Another interviewee, working on a particularly complex project, highlighted the value to their onboarding of reviewing (with the project manager/team leader as well as by themselves) the short- and long-term plans for the team deliverables and work schedules. They noted that it was useful to have documentation of the plans available for referral during onboarding. They explained that they reviewed the plans frequently at the start of onboarding, but less and less as onboarding progressed.

The importance of gaining a solid *product overview* during onboarding was emphasized by a number of interviewees. One participant described how a video presenting an overview of the product's features was particularly useful to refer to during onboarding. During onboarding, the usefulness of developing an understanding of the product's value to the client (i.e., its business value) was noted by several interviewees. They explained that this provided a context and meaning for the development work.

The importance of *team socializing* to the onboarding process was stressed by most interviewees. This was described as times where the team met socially rather than working together. It included organization-wide social gatherings, team celebrations (e.g., going out to lunch to celebrate a project success, sharing a cake for a team member's birthday), as well as informal gatherings (eating lunch together in the lunch room) and 'watercooler-type' interactions. Interviewees stressed that socializing was key to gaining mutual trust and respect during onboarding. One interviewee mentioned that it helped to get acceptance if the team "understands me as a person, not just a developer".

Being assigned a *simple task* at the start of onboarding was front of mind as valuable to onboarding for three of the interviewees. They described the value in undertaking the simple task being the opportunity to apply and contextualize their recent learning. It was also seen as an opportunity to find out what they "didn't know they didn't know" and address this. The importance of the low impact of making a mistake in a simple task or the low impact of being slow to finish the simple task was noted by one interviewee. A quick success and avoiding looking stupid were also described as part of the onboarding value of doing a simple task. In contrast, another (experienced and confident) interviewee noted that they found being "thrown in the deep-end" on a complex task useful during onboarding because it was an opportunity to prove themselves to the team as a step towards gaining their trust and respect.

There were a number of techniques that interviewees described as valuable for improving their work-ready skills and capabilities during onboarding. The most commonly mentioned activity was attendance at an organized *training course* related to the key technology used at the workplace.

Most participants referred to workshop-style short courses run by third parties, although some participants included formal online courses (such as offered by Coursera or EdX).

Such courses were viewed as an opportunity for extended learning from an expert (and possible certification) that could then be contextualized to their team environment. A *course on Agile* approaches to software development was highlighted as critical to their onboarding by one interviewee working in a team using an Agile way of working. These courses were viewed as being organized and paid for by the organization. In contrast, *self-learning* was seen as the initiative of the onboarder and also identified as important to their onboarding by several participants. Reading books, doing online tutorials and attending local Meetup groups were all referred to as self-learning. Attending a *conference* was an important aspect of onboarding for one interviewee. It was seen as an opportunity to learn from others, make contact with experts, identify new resources and network with the community of practice. All aspects were seen as useful for improving capabilities and skills during onboarding. It was also useful for relationship building if other team members attended the conference.

A few interviewees new to their organization mentioned that they attended a formal *Induction* course and that this was valuable for their team onboarding. They were introduced to policies, organizational procedures and some practices that influenced their teamwork. One interviewee worked for short periods of time at different departments and branches of the organization as an orientation to the company and found the information and knowledge gained from these experiences useful to get onboarded to a team.

### 3) Working with Artefacts

*Internal documentation* was identified by most interviewees as important during onboarding to help them to understand software development artefacts such as existing code design decisions, tests, work standards, database structures and the setting up and the use of development tools locally. The internal documentation onboarding category referred to information stored in a variety of places including workflow tools such as Jira, a physical work board (e.g., user story board), comments in the code, information in the code repository and electronic documentation and manuals. Other local documentation specifically mentioned by interviewees as useful for onboarding are the use of a knowledge database, project plans, the existing code base, product information, checklists, and a location map.

*Knowledge databases* are a particular type of local documentation that were singled out by a number of interviewees as contributing significantly to their onboarding. This was typically a wiki or wiki-style database that contains substantial organizational and team knowledge. The notable characteristics mentioned are that it is quick to access, accessible from anywhere, easily searchable, and is up-to-date. One interviewee noted that it gave a sense of progress when they were able to contribute to it during their onboarding and helped them to gain the team's trust. The knowledge database was explained as being useful to get technical information about coding, testing, requirements, work procedures, development tools, and security, as well as organization-wide information such as policies and organizational structure.

Access to a central code repository was also important to onboarding for most interviewees. This allowed them to *review others' code* and gain an understanding of the existing code base design, coding standards and norms, as well as learning new techniques. A couple of interviewees also noted the onboarding value of attending team code reviews.

The usefulness during onboarding of having *checklists* as memory aids (e.g., continuous integration steps, review questions) was emphasized in one interview. Related to this, another interviewee said that having a *location map* of where people in the organization were situated, with their name, role and area of expertise, was invaluable during early onboarding.

### B. Mapping Onboarding Techniques to Onboarding Goals (RQ3)

This subsection describes findings from the aggregated RG results. Table III and Figure 2 present the major contributions (H or VH) of each onboarding technique in Table II, for each onboarding goal in Table I.

These findings in Fig. 2 and Table III can be used to guide an organisation in the selection of onboarding techniques to include in an onboarding program. For example, it can be seen that Mentoring and Peer Support are perceived as having high contributions to *all* onboarding goals and so are likely to be key techniques for any onboarding program, with potentially high payback. On the other hand, their inclusion may need to be balanced against the cost-benefit associated with their implementation. For example, mentor training and freeing up the mentor's time may not be worth it for a particular new team member who is familiar with the organisation, technology and project domain. All of the nine onboarding techniques on the right-hand-side of Fig. 2 have a perceived high contribution to six or more onboarding goals. This contribution profile may place them high on a list of techniques to consider. That is not to say that onboarding techniques on the left-hand-side of Fig. 2 are not worth considering. They may also be important for achieving specific goals that are particular to a given onboarding situation. For example, a new team member who is also new to the organisation may benefit from developing a clearer understanding of company culture during onboarding and so an Induction Course could be a high priority in this case.

These results could be useful for practitioners to design a personalized onboarding program for a given onboarder's characteristics and team situation. For example, it could reasonably be expected that the onboarding needs, and corresponding onboarding goals, would be different for a new graduate just starting with an organization and team, and a new experienced team member swapping from another team in the organization: a set of onboarding goals could be selected that are appropriate to their particular contexts. Based on Table II and Fig.2, a set of onboarding techniques could then be selected that should achieve the onboarding goals that suit those contexts. An example is described in section V.

### C. Duration of Onboarding (RQ4)

How long the onboarding process could be expected to take is of interest to the team, the new team member and management. Understanding this can be used for planning as well as setting and aligning expectations. The following quote from an interviewee exemplifies the view of an experienced developer:

*I started to feel confident with my work after the first month I had worked here, and an extra month to have full control of what I am doing.*

For a new team member with little or no work experience interviewees reported that it required generally one or more months extra to feel onboarded. The following quote is typical:

TABLE III. ACHIEVING THE DESIRED ONBOARDING GOALS

| Goal | Very High contribution | High contribution |
|---|---|---|
| G1. Understanding team norms | Peer Support, Team Leader Assistance | Mentoring, Induction, Training, Checklists, Team Socialising, Stand Ups, Pair Programming, Simple Task, Other Teams, Electronic Communication, Set Expectations, Review Plan, |
| G2. Understanding company culture | Induction | Mentoring, Training, Peer Support, Team Socialising, Pair Programming, Electronic Communication, Set Expectations, Review Plan, |
| G3. Knowing the responsibilities, expertise and authority of other team members | Agile Course, Team Leader Assistance, Review Plan, | Mentoring, Training, Peer Support, Team Socialising, Stand Ups, Other Teams, Electronic Communication, |
| G4. Understand other's expectations of your own role's responsibilities | Agile Course, Team Leader Assistance, Review Plan, | Mentoring, Training, Peer Support, Internal Documentation, Checklists, Pair Programming, Electronic Communication, |
| G5. Understand what work to do and when | Agile Course, Team Leader Assistance, Checklists, Other Teams, Review Plan, | Mentoring, Peer Support, Stand Ups, Pair Programming, Set Expectations, |
| G6. Understand the project structure and aims and the implications | Team Leader Assistance, Product Overview, Review Plan, | Mentoring, Training, Peer Support, Internal Documentation, Stand Ups, Pair Programming, Set Expectations, |
| G7. Understand how to code and test to the team's expectations | Mentoring, Training, Team Leader Assistance, Review Code, Simple Task, | Peer Support, Internal Documentation, Online communities, Pair Programming, Electronic Communication, Self-Learning, Conference, |
| G8. Understand and meet the team's standards of work quality | Agile Course | Mentoring, Internal Documentation, Stand Ups, Pair Programming, Other Teams, |
| G9. Understand and show the agile mind set | Agile Course, Team Leader Assistance. | Mentoring, Training, Peer Support, Internal Documentation, Review Code, Simple Task, Electronic Communication, Conference, |
| G10. Know how to use Agile artefacts and techniques that are part of the team's software development process | Agile Course, Other Teams, Knowledge Database | Mentoring, Training, Peer Support, Internal Documentation, Online communities, Electronic Communication, |
| G11. Understand the project domain knowledge and terminology | Product Overview; Conference; Knowledge Database | Mentoring, Training, Peer support, Internal Documentation; Pair programming, Simple task, Other teams, Electronic communications |

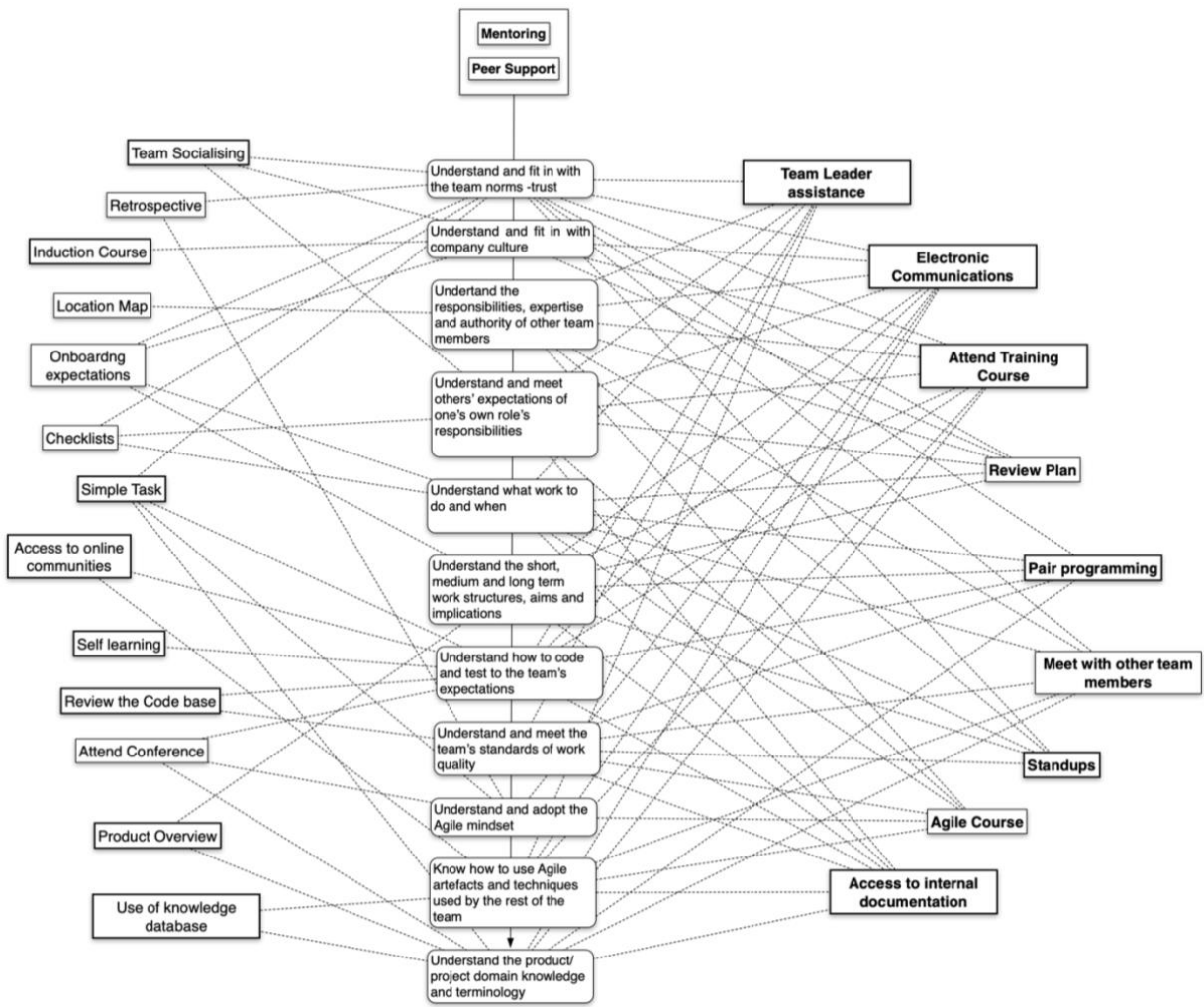

Fig. 2. Mapping onboarding techniques with a high or very high contribution to different onboarding goals

*It was hard for me the learn the languages and techniques at the very beginning and it took me three weeks to learn. After the training session, I was able to do some easy tasks... After around another one and a half month, I was capable of finishing regular tasks, but still needed some help from others.*

While these developers emphasized work tasks, another graduate developer also mentioned the time to feel accepted by the team and understand the process adopted by the team:

*I can now [4 months in] finish my work by myself, but I do not think I am already part of the team. I still get confused with some process during development and still cannot acquaint [sic] everyone in my team.*

The front-of-mind perceptions of both experienced and inexperienced interviewees emphasized task capability as the main indicator of team onboarding ending. Only a few mentioned understanding the team process and feeling like part of the team as expected outomes. This may indicate a lack of clarity in the onboarding goals. Overall, the perception of interviewees was that onboarding was completed in 2-6 months for them, but this needs further research.

## V. IMPLICATIONS FOR IMPLEMENTING ONBOARDING

### A. Onboarding Techniques

The results in Table II provide an informed starting point as a checklist for practitioners to select a set of onboarding techniques to consider including in their own onboarding. This may broaden their repertoire of onboarding techniques they consider explicitly. Furthermore, the organization can better understand the onboarder perspective of the value and challenges of these onboarding techniques from the detailed discussion of each technique in Section IV. Taking these insights into account could help the organization implement the techniques successfully in practice. For example, it was clear from the interviewees that mentoring is viewed as a key onboarding practice, providing information, advice and acting as a confidant to the onboarder.

It is important for an organization to recognize the effort needed to implement these, however, if the onboarding is to be effective. The likely organizational efforts and responsibilities for implementing onboarding techniques can be grouped into the categories shown in Table IV, and are discussed in what follows.

*1) Socialization opportunities*

Interviewees considered involvement in team social events as a significant factor in developing relationships and team trust. The first row in Table IV emphasises that the organization will need to take some responsibility for organizing socialization opportunities that can be attended by the onboarder. This may involve budgeting for time and money to support such activities, as well as encouraging this as an organizational norm.

*2) Access to high quality knowledge artefacts*

The second row of table IV focuses on the responsibility of the organization to support the information-seeking of an onboarder (and the related organizational information sharing) as they improve their knowledge of how they should work, what they should work on and gain the competencies and skills to do the work to the team's expectations. Effective and efficient onboarding relies on the onboarder having timely and low-effort access to relevant information from these sources, as well as the quality of the information. The onboarder needs to be able to access the onboarding-relevant internal electronic documents and knowledge bases (e.g. security clearance, logons, instructions), while still being protected from inadvertent modification of the information.

The onboarder may also need some training in searching techniques and evaluation of information credibility. Ideally the text-based information should be available from wherever the onboarder works and at all times, so it is available at the time of need. Several interviewees reported that information in the knowledge database was sometimes outdated or that they found errors in the internal documentation, highlighting the need for such information to be accurate and up-to-date. It may be useful to provide the onboarder with an artefact that shows what information is available from what sources (e.g., 'yellow pages') if this is particularly complex.

*3) Access to formal training*

The organization may need to organize for the onboarder to attend training courses to meet their anticipated and discovered skill gaps. This may require the organization to set aside some funds to pay for the training as well as schedule time for the onboarder to attend the courses and organize course registration.

*4) Proactive feedback and knowledge sharing*

For onboarding techniques that rely on organizational insiders for information, advice or feedback, there may be a need to train onboarding-related people to anticipate the onboarder's information and feedback needs and have the skill and motivation to address them. For example, interviewees report that it was useful to get feedback on onboarding progress and work quality during onboarding. The organization must also acknowledge the value of the time and effort of those around the onboarder in sharing knowledge and feedback. For both the onboarder and onboarding-related people the expectations of availability should be aligned. A location map about who has expertise in what areas may be useful and this will require ongoing effort to keep up-to-date.

TABLE IV. ONBOARDING ORGANIZATIONAL EFFORT

|   | Category | Related Onboarding Techniques |
|---|---|---|
| 1. | *Socialization opportunities* | -Team socializing<br>-Meet other teams |
| 2. | *Access to high quality knowledge artefacts* | -Knowledge database<br>-Internal Documentation<br>-Review code base<br>-Review plans<br>-Online communities<br>-Location maps<br>-Checklists |
| 3. | *Access to formal training* | -Training course<br>-Agile course<br>-Conference Attendance |
| 4. | *Proactive feedback and knowledge sharing* | -Mentoring<br>-Team Leader<br>-Peer support<br>-Set expectations |
| 5. | *Provide psychological safety to experiment and learn* | -Simple task<br>-Pair programming<br>-Retrospectives<br>-Stand ups |

Taking this anxiety away provided a better environment for learning and trying things out. This needs the organization and team to make the effort to explicitly promote a mindset and values that focus on learning from mistakes and constant improvement.

*1) Provide psychological safety*

Most interviewees noted that their onboarding was more effective if they felt that they would not be punished or blamed for making a mistake or being slower than others in the team to produce work.

*B. Contextualized Onboarding Design*

Designing a contextualized or personalized onboarding program relies on selecting a set of onboarding goals from Table I that can be customized to suit the needs of the onboarder and other specific project and team contextual factors. Given a set of goals, the set of onboarding techniques with high contributions to these goals can then be selected, based on Table III and Fig. 2. Consideration of team onboarding scenarios mentioned by interviewees provides a basis for the onboarder characteristics that could be important. Some example scenarios are: a new graduate employee; a new employee experienced in the technology and product but not in an Agile way of working; a temporary team formed from existing employees for a defined scope of work; a person from a different workstream onboarding to the new team to replace a team member leaving; a new team leader from elsewhere in the organization but with no previous team leadership experience; a new team formed from long-term and new employees.

These scenarios suggest examples of different onboarder characteristics that could influence the selection of different onboarding goal sets: work experience (e.g., new graduate or seasoned veteran), familiarity with the organizational culture (e.g., new or existing long term employee); previous work relationships with the new team members (e.g., trusted by team members because of previous teamwork); role in the new team (e.g., intern or team leader); level of related technical and product domain experience (e.g., new to the product's technology and tools or the knowledge domain of the product); experience with the team's agile way of working (e.g., new to Agile or internalized Agile values and experience with different Agile ways of working); the life expectancy of the team (e.g., temporary or long term team); the resources available (e.g., small versus large organization).

As an example of selecting a goal set and the corresponding onboarding techniques for a specific scenario, consider the case of an existing employee joining a team. They have been working in the same role with a different team using an Agile way of working, but they are unfamiliar with the product and one of the development tools and have not worked with anyone from this team before. A suitable set of team onboarding goals to focus on for this new team member could be: G1 (understand team norms), G3 (understand others' responsibilities, expertise and authority), G6 (understand planned work structures and aims) G10 (understand the tools and techniques), G11 (understand the product domain). The corresponding onboarding techniques to focus on implementing would therefore be: Team socializing, Access to online communities, Access to internal documentation, Product overview, Review plans, and Standups. A Training course on the unfamiliar tool may also be effective.

## VI. CONCLUSION

This study first investigated the 'how' and 'why' of onboarding a new team member in the context of Agile software development teams. A set of eleven categories of onboarding goals (the 'why') was synthesized from a review of current literature, answering RQ1 (Table I). To answer RQ2 an interview survey involving eleven participants from eight organizations was conducted. A list of twenty-four diverse onboarding practices and techniques (the 'what') was discovered from a synthesis of the interview data (Table II). The interview data also suggested that an organization can expect to support the onboarding of a new team member for around 3 months (RQ4). As well as providing a useful list of diverse onboarding practices to consider, the findings from the interview data provided a number of insights about the value and challenges of the onboarding techniques. These, together with organizational efforts for implementation, may help organizations select from a wider repertoire of onboarding techniques than they were previously aware of, as well as provide guidance for their successful implementation.

The second part of the research considered how to select onboarding techniques for a given situation. A Repertory Grid Technique was used to collect and aggregate interviewees' perceptions of how much each onboarding technique contributed to the achievement of each onboarding goal (RQ3). It is expected that understanding these relationships will help to guide organizations to design successful onboarding programs by providing an evidence-based mechanism for selecting techniques to be included that should achieve a particular goal profile. This goal profile could be selected to suit the context of a specific onboarding situation, an area of planned future research.

## VII. LIMITATIONS AND THREATS TO VALIDITY

Construct validity relates to the alignment of what was investigated to what the researchers had in mind. To avoid ambiguities and misunderstandings, definitions of the study goals and of the key term "onboarding" were included in the interview guide and clarified with interviewees. A pilot interview was conducted to test the study protocol and helped to improve the clarity of the questions and eliminate those seen as leading participants. Interviewers were sensitive to avoiding influencing the interviewees' answers and were silent during the repertory grid development.

Some generalizability (external validity) can be claimed for the context of developers and testers in Agile software development teams. Since the interviewees were reflective of diverse organizational contexts and individual experiences, we expect new Agile team members would have similar onboarding goals and benefit from the related onboarding techniques. Further work is needed to evaluate the effectiveness of using the technique-goal mapping for personalizing onboarding design.

The steps in the research process are described in sufficient detail to ensure reliability of the data collection and analysis. The coding and categorization of goals and techniques has an unavoidable degree of subjectivity. To mitigate this threat, two researchers independently coded and categorized the data and aligned the results. The protocol design was reviewed by a third researcher and informed by the pilot. The code set, categories and results were also reviewed by representatives from some of the participating organizations.

It is unlikely that the list of onboarding techniques identified by the participants in this study is exhaustive. As Kelly suggests, people's cognitive models of phenomena are "subject to as great a variety of constructions as our wits will enable us to contrive" [2]. In addition, it is inevitable that new techniques will be developed as new ideas are experimented with and refined.